\documentclass[prc,showpacs,twoside,preprintnumbers,amsmath,amssymb]{revtex4-1}

\usepackage{graphicx,color}
\usepackage{dcolumn}
\usepackage{bm}

\begin{document}

\title{Microscopic Description of Spherical to $\gamma$-soft Shape Transitions in Ba and Xe Nuclei}

\author{Z. P. Li$^{1,2}$}
\author{T. Nik\v{s}i\'{c}$^{2}$}
\author{D. Vretenar$^{2}$}
\author{J. Meng$^{1,3}$}

\affiliation{$^{1}$State Key Laboratory of Nuclear Physics and Technology, School of Physics,
Peking University, Beijing 100871, China}
\affiliation{$^{2}$Physics Department, Faculty of Science, University of Zagreb,
10000 Zagreb, Croatia}
\affiliation{$^{3}$School of Physics and Nuclear Energy Engineering,
Beihang University, Beijing 100191, China}

%

\date{\today}

\begin{abstract}
The rapid transition between spherical and $\gamma$-soft shapes in Ba and Xe nuclei in the
mass region $A \geq 130$ is analyzed using excitation spectra and collective wave functions
obtained by diagonalization of a five-dimensional Hamiltonian for quadrupole vibrational and
rotational degrees of freedom, with parameters determined by constrained self-consistent
relativistic mean-field calculations for triaxial shapes. The results reproduce the characteristic
evolution of excitation spectra and E2 transition probabilities and, in general, a good
agreement with available data is obtained. The calculated spectra display fingerprints
of a second-order shape phase transition that can approximately be described by analytic 
solutions corresponding to the E(5) dynamical symmetry.

\end{abstract}


\maketitle
\section{Introduction}
Atomic nuclei display a variety of different equilibrium shapes -- spherical,
axially deformed, shapes that are soft with respect to triaxial deformations.
In most cases the transitions between different shapes in isotopic or
isotonic sequences are gradual, and reflect the underlying modifications
of single-nucleon shell structure and interactions between valence nucleons.
In a number of cases, however, with the addition or subtraction of only few
nucleons one finds experimental signatures of abrupt changes in observables
that characterize ground-state nuclear shapes. In the last decade the
concept of quantum phase transitions (QPT) has successfully been applied
and investigated, both experimentally and theoretically, in equilibrium shape
changes of nuclei \cite{Iac.03,JC.05,Rick.06,Rick.07,CJ.08,Bon.08}.

The two best studied classes of nuclear shape phase transitions correspond
to a second-order QPT between spherical and $\gamma$-soft shapes
\cite{FI.00,CZ.00}, and a first-order QPT between spherical and
axially deformed shapes \cite{FI.01,CZ.01}. Most theoretical studies of QPT in 
nuclei have been based on model specific Hamiltonians that by
construction describe shape changes, i.e. a phase transition is accessed by 
variation of a control parameter explicitly built into the Hamiltonian.  For instance, 
a QPT can be described in the geometric framework in terms of a Bohr Hamiltonian 
for shape variables, and related to the concept of critical symmetries that provide
parameter independent predictions for excitation spectra and electric quadrupole (E2)
 transition rates for nuclei at the phase transition point. Analytic solutions of the
eigenvalue problem at the critical point are associated with zeros of
special functions. Alternatively, in the algebraic 
approach different shapes coincide with particular dynamic symmetries of some
algebraic structure, and a QPT may occur when these symmetries are broken in
a specific way. An example of the latter approach is the framework of the 
interacting-boson model (IBM) \cite{IA.87} and its various extensions.  

More recently a number of studies appeared that attempt a fully microscopic
description of shape QPT starting from nucleonic degrees of
freedom \cite{Meng.05,SG.05,FBL.06,Ni.07,RS.07,RE.08,RRS.08,Li.09,Li.09a}.
An approach in terms of explicit nucleonic degrees of freedom is particularly
important because the physical control parameter in a nuclear QPT is the
actual number of nucleons. In a series of 
studies \cite{Ni.07,Li.09,Li.09a} we investigated nuclear QPT in the region
$Z = 60$, $62$, $64$ with $N \approx 90$, using a microscopic approach
based on constrained self-consistent relativistic mean-field calculations
of potential energy surfaces. Since a quantitative analysis of QPT must go beyond
a simple mean-field level, i.e. one must be able to calculate ratios of excitation
energies and electromagnetic transition rates, in Ref.~\cite{Ni.07} the generator
coordinate method (GCM) was used to perform configuration
mixing of angular-momentum and particle-number projected relativistic
wave functions restricted to axial symmetry. This approach was extended
in \cite{Li.09,Li.09a}, where collective excitation spectra
and transition probabilities were calculated starting from a five-dimensional
Hamiltonian for quadrupole vibrational and rotational degrees of
freedom, with parameters determined by constrained mean-field calculations for
triaxial shapes, i.e. including both $\beta$ and $\gamma$ deformations.
The results reproduced available data, and have shown that there is an abrupt
change of structure at $N=90$ that can approximately be characterized by the
X(5) analytic solution \cite{FI.01} at the critical point of the first-order quantum phase
transition between spherical and axially deformed shapes.  As a function of the
physical control parameter -- the number of nucleons, energy gaps between the
ground state and the excited vibrational states with zero angular momentum,
isomer shifts, and the monopole transition strength, exhibit sharp discontinuities at
neutron number $N=90$, characteristic of a first-order quantum phase transition.

Are the remarkable results for X(5)-like shape phase transitions in the mass region
$A \approx 150$ \cite{Ni.07,Li.09,Li.09a} somewhat accidental? Can the same 
universal energy density functional describe other types of shape phase transitions, 
and in different regions of the chart of nuclides? To answer these questions
we apply the same model to the study of $\gamma$-soft shapes in Ba and Xe nuclei in the
mass $A \geq 130$ region, where evidence has been reported for a second-order QPT
between spherical and $\gamma$-soft shapes. This is a phase
transition in one degree of freedom -- the axial deformation
$\beta$ and, in the IBM language, represents a transition between the $U(5)$ and
$O(6)$ dynamical symmetries in the limit of large boson number.
The critical point of phase transition can also be related to a dynamical
symmetry: in this case E(5) \cite{FI.00}, and the experimental realization of this
critical-point symmetry was first identified in the spectrum of $^{134}$Ba \cite{CZ.00}.

The model is based on constrained self-consistent relativistic mean-field
calculations for triaxial shapes, i.e. including both $\beta$ and $\gamma$ deformations.
The resulting self-consistent solutions: single-particle
wave functions, occupation probabilities, and quasiparticle energies that correspond to
each point on the binding energy surface, are used to calculate the parameters that determine
the collective Hamiltonian: three mass parameters, three moments of inertia, and the zero-point
energy corrections, as functions of the deformations $\beta$ and  $\gamma$ \cite{Ni.09}.
The diagonalization of the Hamiltonian yields the excitation energies and collective wave
functions that are used to calculate observables. No attempt is made to tune the model or the
parameters to this particular type of shape transition or mass region. As in our previous studies
of X(5)-like transitions  \cite{Ni.07,Li.09,Li.09a}, the relativistic functional PC-F1 (point-coupling
Lagrangian)~\cite{BMM.02} is used in the particle-hole channel, and a density-independent
$\delta$-force is the effective interaction in the particle-particle channel, with pairing correlations
treated in the BCS approximation.
\section{Shape transition in $\bm{Ba}$ and $\bm{Xe}$ isotopes}
Our analysis starts with the self-consistent relativistic mean-field plus
BCS calculation of quadrupole energy surfaces of Ba and Xe isotopes.
The map of the energy surface as function of the quadrupole
deformation is obtained by imposing constraints on the axial and
triaxial mass quadrupole moments. The method of quadratic constraints
uses an unrestricted variation of the function
\begin{equation}
\langle H\rangle
   +\sum_{\mu=0,2}{C_{2\mu}\left(\langle \hat{Q}_{2\mu}  \rangle - q_{2\mu}  \right)^2} \; ,
\label{constr}
\end{equation}
where $\langle H\rangle$ is the total energy, and
$\langle \hat{Q}_{2\mu}\rangle$ denotes the expectation value of the mass quadrupole
operators:
\begin{equation}
\hat{Q}_{20}=2z^2-x^2-y^2 \quad \textnormal{and}\quad \hat{Q}_{22}=x^2-y^2 \;.
\end{equation}
$q_{2\mu}$ is the constrained value of the multipole moment,
and $C_{2\mu}$ the corresponding stiffness constant~\cite{RS.80}.

\begin{figure}[htb]
\hspace{-2cm}
\includegraphics[scale=0.7]{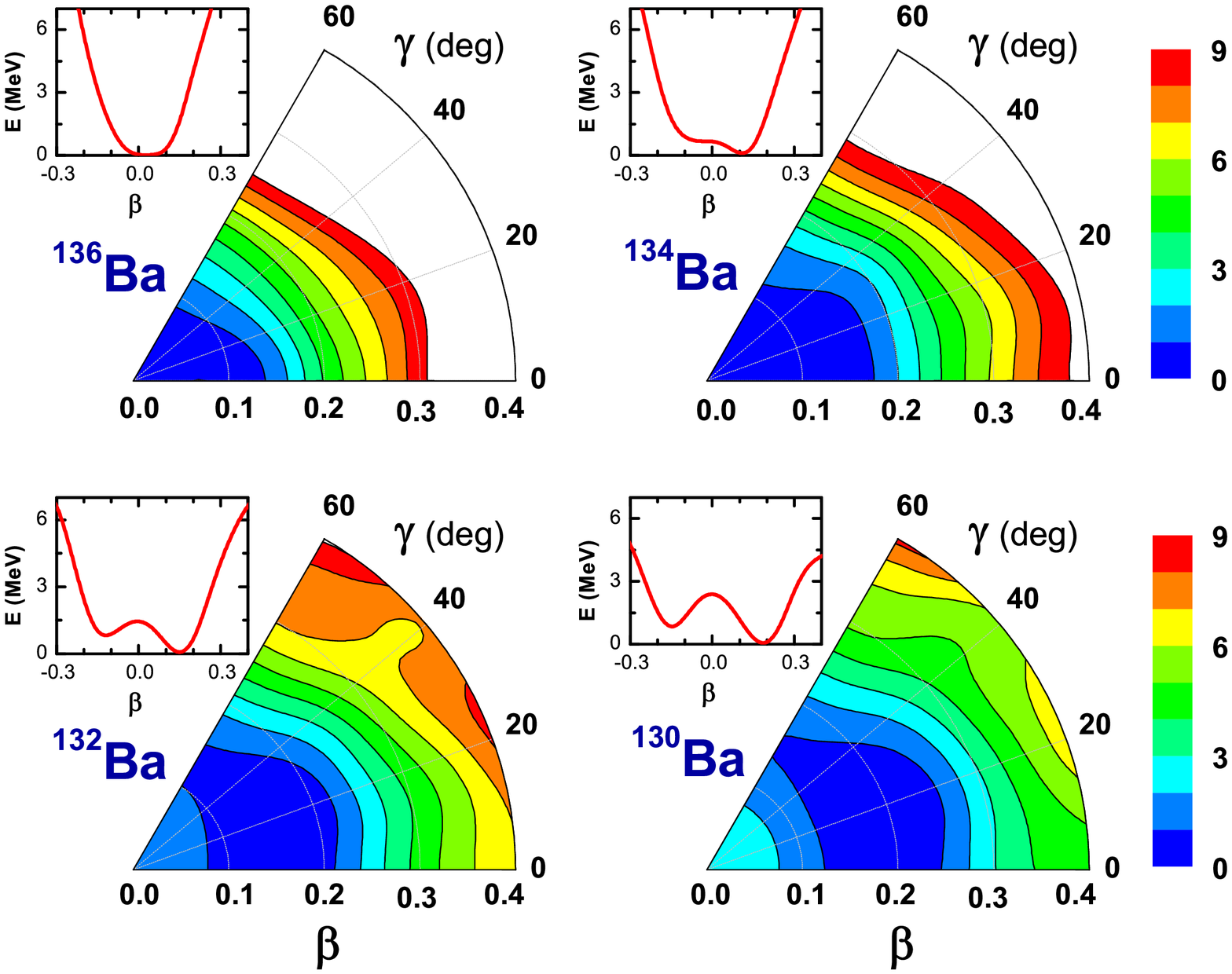}
\vspace{-2cm}
\caption{\label{PES-Ba}(Color online) Self-consistent RMF+BCS triaxial quadrupole
binding energy maps of the even-even $^{136-130}$Ba isotopes
in the $\beta - \gamma$ plane ($0\le \gamma\le 60^0$).
All energies are normalized with respect to
the binding energy of the absolute minimum, the contours join points
on the surface with the same energy (in MeV). For each nucleus in the
corresponding inset we plot the axial projection of the binding energy
for oblate (negative $\beta$) and prolate (positive $\beta$) deformations.}
\end{figure}

\begin{figure}[htb]
\hspace{-2cm}
\includegraphics[scale=0.7]{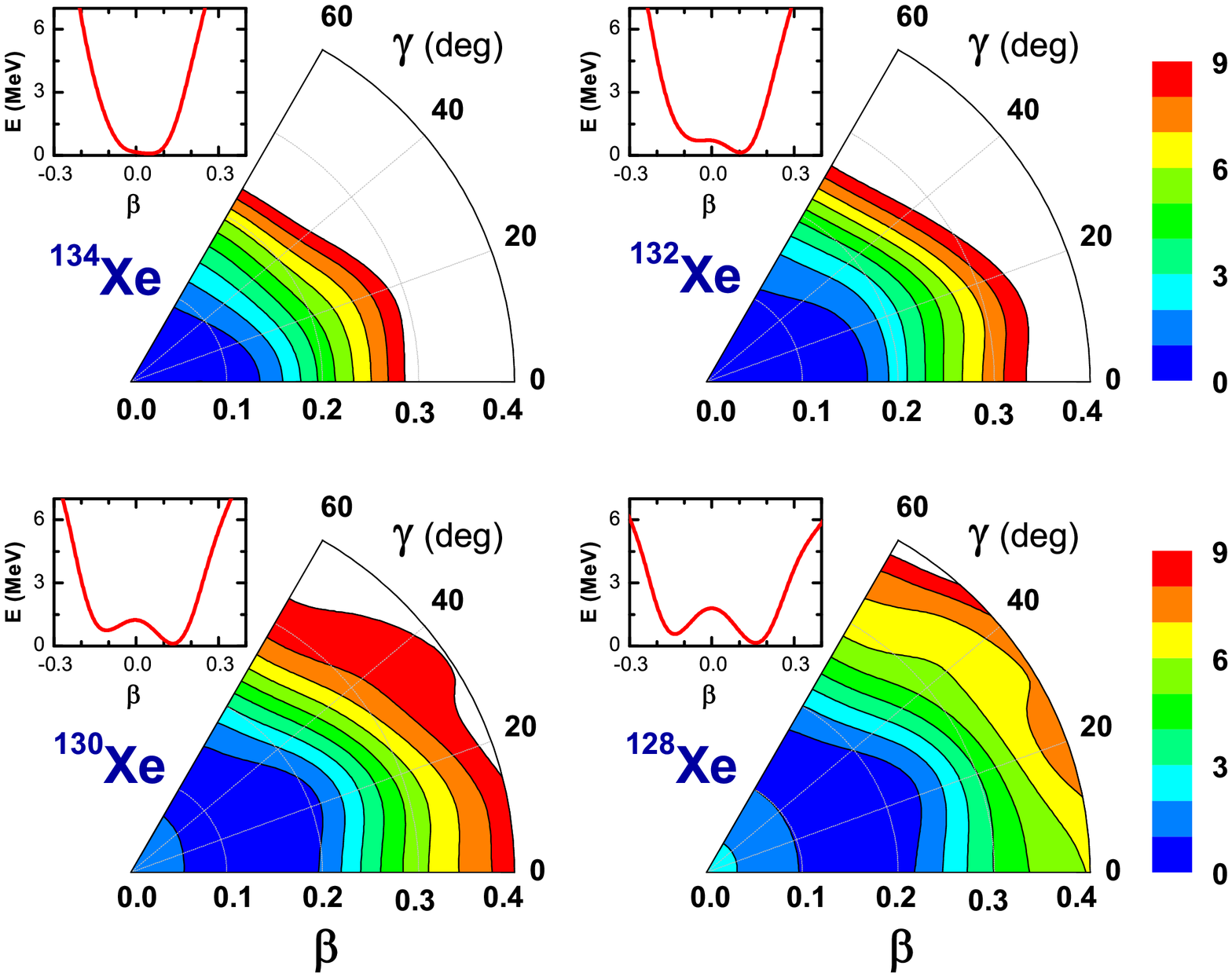}
\vspace{-2cm}
\caption{ \label{PES-Xe} (Color online)
Same as described in the caption to Fig.~\ref{PES-Ba} but for the isotopes
$^{134-128}$Xe.}
\end{figure}

In Fig.~\ref{PES-Ba} we display the RMF+BCS triaxial quadrupole binding
energy maps of the even-even $^{136-130}$Ba in the $\beta - \gamma$ plane
 ($0\le \gamma\le 60^0$). All energies are normalized with respect to
the binding energy of the absolute minimum, and the contours join points
on the surface with the same energy (in MeV). For each nucleus in the
corresponding inset we plot the axial projection of the binding energy
for oblate (negative $\beta$) and prolate
(positive $\beta$) deformations. The corresponding
energy maps of the even-even $^{134-128}$Xe isotopes are shown in
Fig.~\ref{PES-Xe}. Both for Ba and Xe, these plots illustrate the rapid
transition from spherical shapes near the $N=82$ closed-shell, to
$\gamma$-soft energy surfaces for lighter isotopes. Starting from almost
perfect spherical shapes in $^{136}$Ba and $^{134}$Xe, the decrease in
neutron number induces quadrupole deformations. In the axial plots we
notice that the potentials become less stiff in $\beta$ and that two minima
develop, one prolate and one oblate, at almost the same energy, and
separated by a low spherical barrier. However, the 3-dimensional energy
maps show that the spherical barriers are concentrated around $\beta \approx 0$
and, therefore, rather than two separate minima, the potentials display continuous
$\gamma$-soft minima that extend from prolate to oblate shapes.

\begin{figure}[htb]
\includegraphics[scale=0.5]{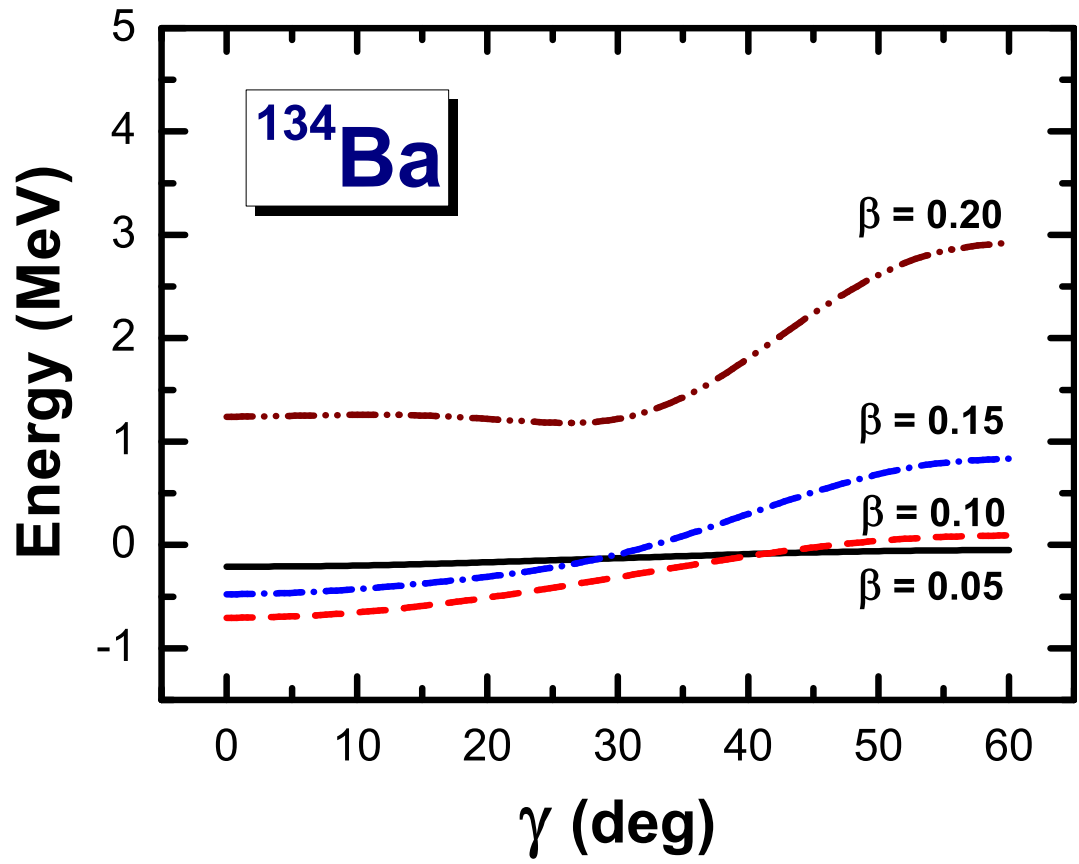}
\caption{ \label{PEC-134Ba} (Color online)
Self-consistent RMF+BCS binding energy curves of the $^{134}$Ba
nucleus, as functions of the deformation parameter $\gamma$,
for four values of the axial deformation $\beta =0.05$, $0.1$, $0.15$, and $0.2$.}
\end{figure}

\begin{figure}[htb]
\includegraphics[scale=0.5]{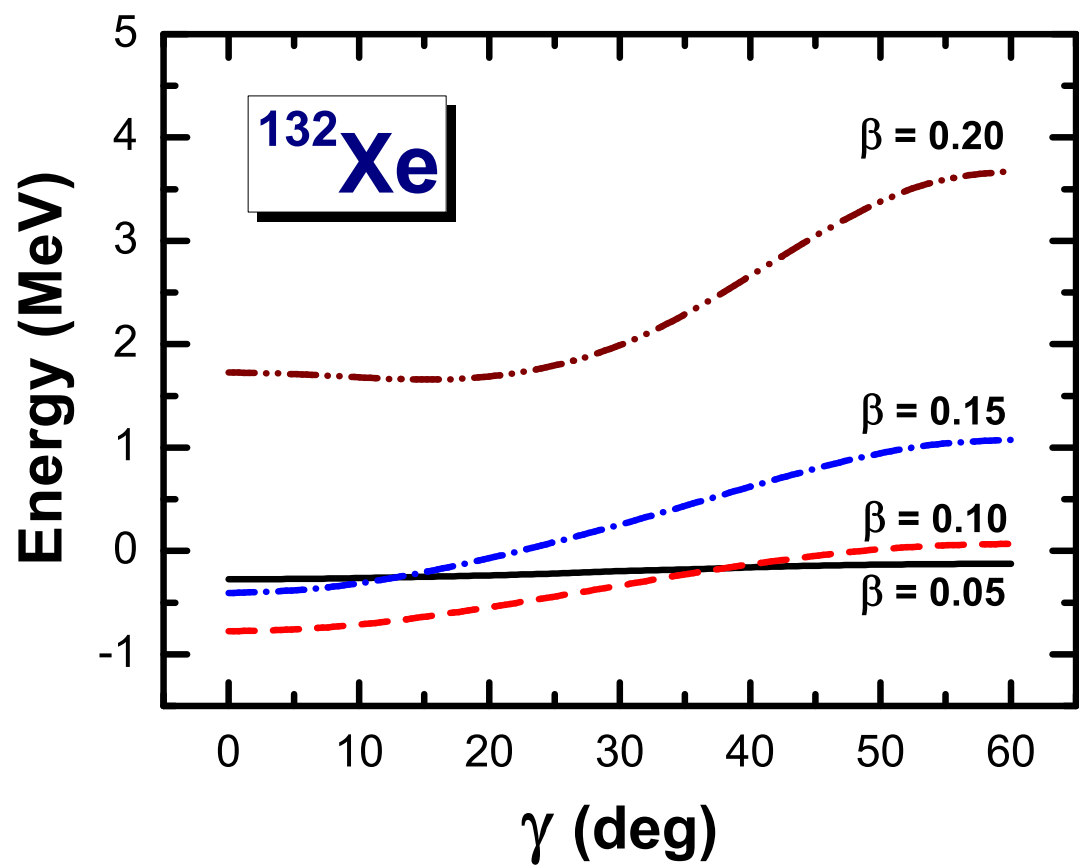}
\caption{\label{PEC-132Xe} (Color online)
Same as described in the caption to Fig.~\ref{PEC-134Ba} but for the nucleus
$^{132}$Xe.}
\end{figure}

Of particular interest in the present analysis are the nuclei which have been identified
as possible candidates for a shape phase transition that can be characterized by the
E(5) dynamical symmetry \cite{FI.00}.  The experimental realization of this critical-point
symmetry, associated with a second order quantum phase transition between spherical
and $\gamma$-soft potential shapes, was first identified in $^{134}$Ba \cite{CZ.00}.
E(5) is the symmetry of a 5-dimensional (intrinsic variables $\beta$ and $\gamma$ and the
three Euler angles) infinite well in the axial deformation variable
$\beta$ ( $V(\beta) = 0$ for $|\beta| \leq \beta_W$, and $V(\beta) = \infty $ for $|\beta| > \beta_W$),
and the potential is completely $\gamma$-independent. The microscopic binding energy
curve $E(\beta)$ of $^{134}$Ba (Fig.~\ref{PES-Ba}) displays a shape that is almost
symmetric with respect to $\beta = 0$. One notices a relatively flat bottom between
$\beta \approx -0.1$ and $\beta \approx 0.1$ (the oblate configuration is only
$\approx 0.5$ MeV above the prolate minimum), and the potential is rather stiff for
$|\beta| > 0.15$. The dependence on the triaxial deformation parameter $\gamma$ is
seen in the corresponding 3-dimensional energy map and, even more clearly, in
Fig.~\ref{PEC-134Ba} where we plot the binding energy curves as functions of $\gamma$
for several values of the axial deformation: $\beta =0.05$, $0.1$, $0.15$, and $0.2$.
In the region of the flat bottom $|\beta| \leq 0.1$ the binding energy of $^{134}$Ba is
indeed almost independent of $\gamma$, and even for somewhat larger deformations
$0.1 \leq |\beta| \leq 0.2$, only a weak dependence on $\gamma$ is predicted by the
calculation based on the PC-F1 functional. A very similar energy surface is
calculated for the isotone $^{132}$Xe (Figs.~\ref{PES-Xe} and \ref{PEC-132Xe}).

In the next step the constrained self-consistent solutions of the relativistic mean-field plus
BCS equations, i.e. the single-particle wave functions, occupation probabilities, and
quasiparticle energies that correspond to each point on the binding energy surfaces of
Figs.~\ref{PES-Ba} and \ref{PES-Xe}, are used to calculate the parameters that determine
the collective Hamiltonian \cite{Ni.09}:
\begin{equation}
\label{hamiltonian-quant}
\hat{H} = \hat{T}_{\textnormal{vib}}+\hat{T}_{\textnormal{rot}}
              +V_{\textnormal{coll}} \; ,
\end{equation}
with the vibrational kinetic energy:
\begin{align}
\hat{T}_{\textnormal{vib}} =&-\frac{\hbar^2}{2\sqrt{wr}}
   \left\{\frac{1}{\beta^4}
   \left[\frac{\partial}{\partial\beta}\sqrt{\frac{r}{w}}\beta^4
   B_{\gamma\gamma} \frac{\partial}{\partial\beta}
   - \frac{\partial}{\partial\beta}\sqrt{\frac{r}{w}}\beta^3
   B_{\beta\gamma}\frac{\partial}{\partial\gamma}
   \right]\right.
   \nonumber \\
   &+\frac{1}{\beta\sin{3\gamma}}\left.\left[
   -\frac{\partial}{\partial\gamma} \sqrt{\frac{r}{w}}\sin{3\gamma}
      B_{\beta \gamma}\frac{\partial}{\partial\beta}
    +\frac{1}{\beta}\frac{\partial}{\partial\gamma} \sqrt{\frac{r}{w}}\sin{3\gamma}
      B_{\beta \beta}\frac{\partial}{\partial\gamma}
   \right]\right\} \; ,
\end{align}
and rotational kinetic energy:
\begin{equation}
\hat{T}_{\textnormal{\textnormal{\textnormal{rot}}}} =
\frac{1}{2}\sum_{k=1}^3{\frac{\hat{J}^2_k}{\mathcal{I}_k}} \; .
\end{equation}
$V_{\textnormal{coll}}$ is the collective potential.
$\hat{J}_k$ denotes the components of the angular momentum in
the body-fixed frame of a nucleus, and the mass parameters
$B_{\beta\beta}$, $B_{\beta\gamma}$, $B_{\gamma\gamma}$, as well as
the moments of inertia $\mathcal{I}_k$, depend on the quadrupole
deformation variables $\beta$ and $\gamma$:
\begin{equation}
\mathcal{I}_k = 4B_k\beta^2\sin^2(\gamma-2k\pi/3) \;.
\end{equation}
Two additional quantities that appear in the expression for the vibrational energy:
$r=B_1B_2B_3$, and $w=B_{\beta\beta}B_{\gamma\gamma}-B_{\beta\gamma}^2 $,
determine the volume element in the collective space. The moments of inertia are
calculated from the Inglis-Belyaev formula:
\begin{equation}
\label{Inglis-Belyaev}
\mathcal{I}_k = \sum_{i,j}{\frac{\left(u_iv_j-v_iu_j \right)^2}{E_i+E_j}
  \langle i |\hat{J}_k | j  \rangle |^2}\quad k=1,2,3,
\end{equation}
where $k$ denotes the axis of rotation, and the summation runs over
the proton and neutron quasiparticle states. The mass parameters associated with
the two quadrupole collective coordinates
$q_0=\langle\hat{Q}_{20}\rangle$ and $q_2=\langle\hat{Q}_{22}\rangle$
are also calculated in the cranking approximation
\begin{equation}
\label{masspar-B}
B_{\mu\nu}(q_0,q_2)=\frac{\hbar^2}{2}
 \left[\mathcal{M}_{(1)}^{-1} \mathcal{M}_{(3)} \mathcal{M}_{(1)}^{-1}\right]_{\mu\nu}\;,
\end{equation}
with
\begin{equation}
\label{masspar-M}
\mathcal{M}_{(n),\mu\nu}(q_0,q_2)=\sum_{i,j}
 {\frac{\left\langle i\right|\hat{Q}_{2\mu}\left| j\right\rangle
 \left\langle j\right|\hat{Q}_{2\nu}\left| i\right\rangle}
 {(E_i+E_j)^n}\left(u_i v_j+ v_i u_j \right)^2}\;.
\end{equation}
Finally, the potential $V_{\textnormal{coll}}$ in the collective Hamiltonian
Eq.~(\ref{hamiltonian-quant}) is obtained by subtracting the zero-point
energy corrections from the total energy that corresponds to the
solution of constrained RMF+BCS equations, at each point on the triaxial
deformation plane.

The Hamiltonian Eq.~(\ref{hamiltonian-quant}) describes quadrupole vibrations,
rotations, and the coupling of these collective modes. The corresponding
eigenvalue problem is solved using an expansion of eigenfunctions in terms
of a complete set of basis functions that depend on the deformation variables $\beta$ and
$\gamma$, and the Euler angles $\phi$, $\theta$ and $\psi$ \cite{Ni.09}. The diagonalization
of the Hamiltonian yields the excitation energies and collective wave functions:

\begin{equation}
\label{wave-coll}
\Psi_\alpha^{IM}(\beta,\gamma,\Omega) =
  \sum_{K\in \Delta I}
           {\psi_{\alpha K}^I(\beta,\gamma)\Phi_{MK}^I(\Omega)}.
\end{equation}
The angular part corresponds to linear combinations of Wigner
functions
\begin{equation}
\label{Wigner}
\Phi_{MK}^I(\Omega)=\sqrt{\frac{2I+1}{16\pi^2(1+\delta_{K0})}}
\left[D_{MK}^{I*}(\Omega)+(-1)^ID_{M-K}^{I*}(\Omega) \right] \; ,
\end{equation}
and the summation in Eq. (\ref{wave-coll}) is over the allowed set  of
the $K$ values:
\begin{equation}
\Delta I = \left\{ \begin{array}{lcl}
   0,2,\dots,I \quad &\textnormal{for} \quad  &I\; \textnormal{mod}\; 2 = 0 \\
   2,4,\dots,I-1 \quad &\textnormal{for} \quad   &I\; \textnormal{mod}\; 2 =1\; .
\end{array} \right .
\end{equation}

\begin{figure}[htb]
\includegraphics[scale=0.5]{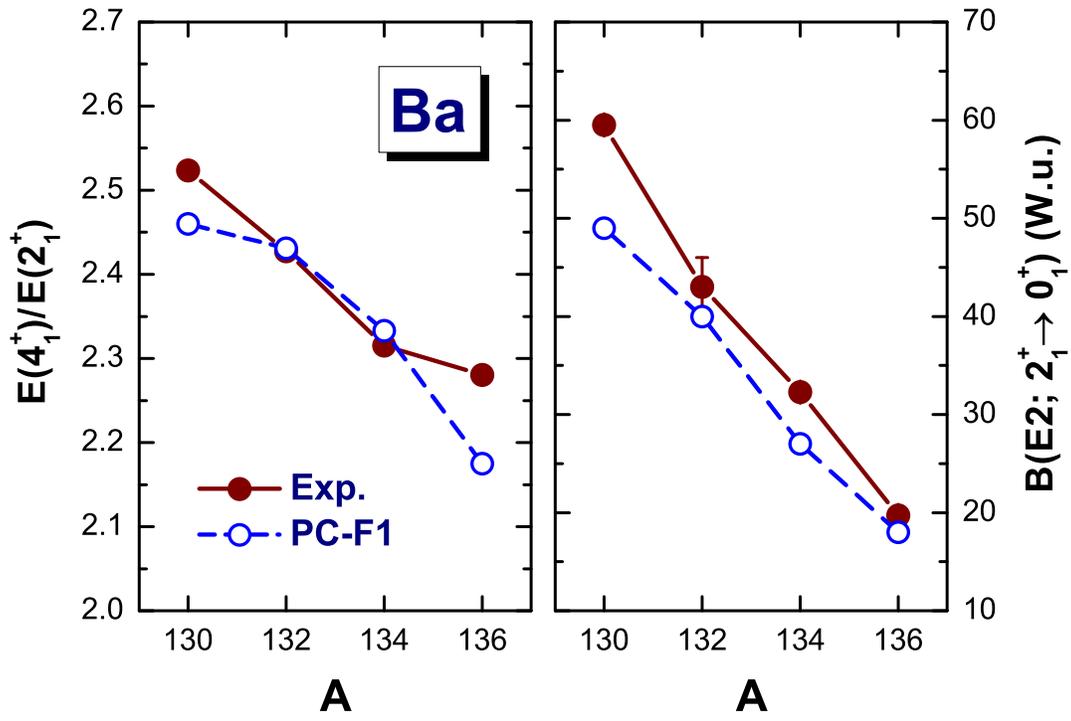}
\caption{\label{R42-Ba} (Color online) Evolution of the characteristic collective observables
$R_{4/2}$ and B(E2; $2^+_1 \to 0^+_1$) (in Weisskopf units) with mass number in
Ba isotopes. The microscopic values calculated with the PC-F1 energy
density functional are shown in comparison with data \cite{NNDC,LBNL}.}
\end{figure}

\begin{figure}[htb]
\includegraphics[scale=0.5]{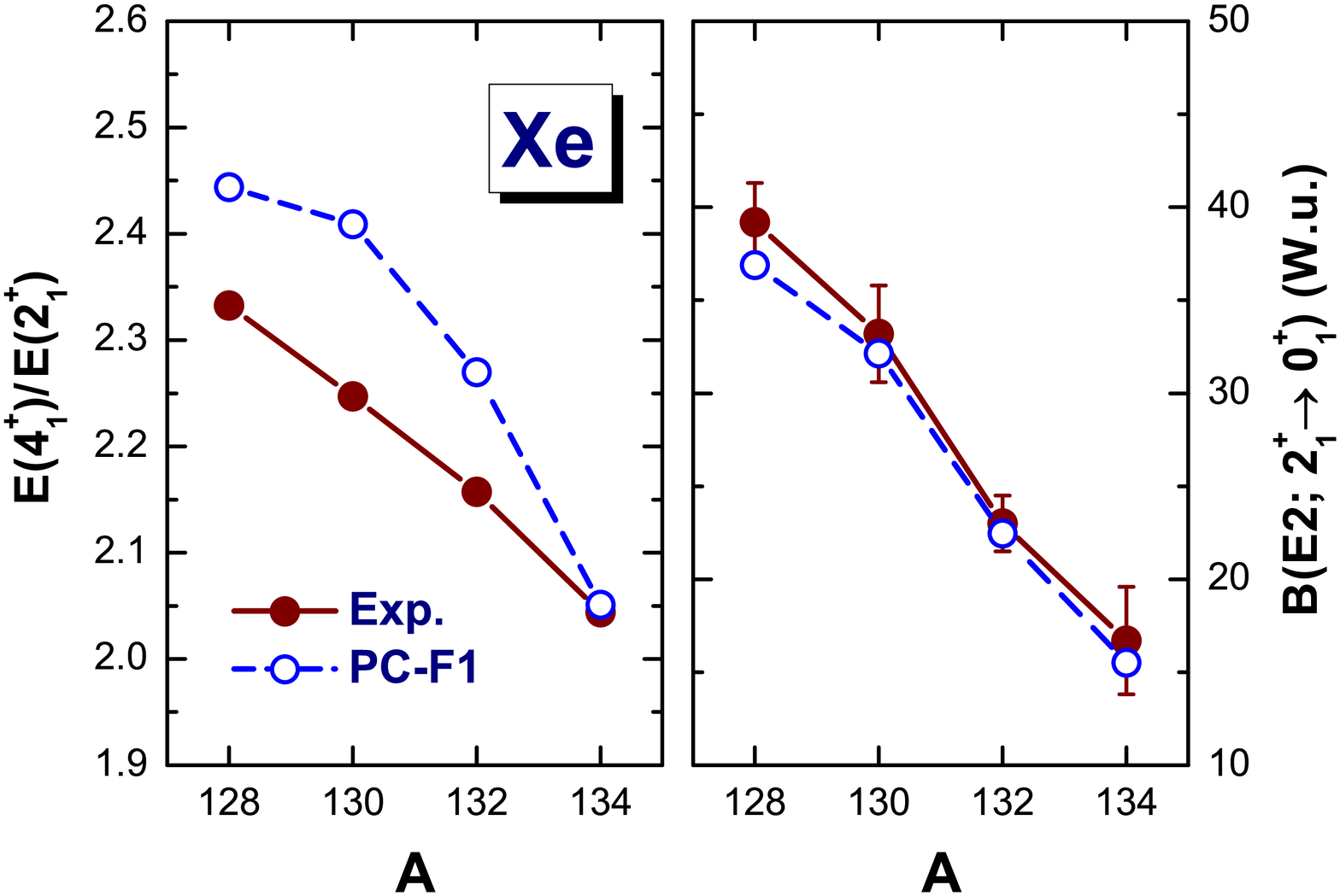}
\caption{\label{R42-Xe} (Color online) Same as described in the caption to Fig.~\ref{R42-Ba} but
for the isotopes $^{128-134}$Xe.}
\end{figure}

In Figs.~\ref{R42-Ba} and \ref{R42-Xe} we plot the isotopic dependence of
two characteristic collective observables: $R_{4/2}$ and B(E2; $2^+_1 \to
0^+_1$) (in Weisskopf units), for Ba and Xe nuclei, respectively. The values
obtained by the diagonalization of the collective Hamiltonian Eq.~(\ref{hamiltonian-quant}),
with parameters determined by the microscopic relativistic energy density functional PC-F1,
are shown in comparison with data \cite{NNDC,LBNL}.  For a transition between
$\gamma$-soft rotors ($O(6)$ dynamical symmetry limit of the interacting boson model)
and spherical vibrators ($U(5)$ dynamical symmetry limit), the ratio between the excitation
energies of the first $4^+$ and $2^+$ states  varies from the value
$R_{4/2} = 2.5$ in the $O(6)$ limit, and $R_{4/2} = 2$ for a spherical vibrator.
For a second-order shape phase transitions between these limits, the parameter-free
E(5)-symmetry prediction at the critical point is $R_{4/2} = 2.2$.
The calculation reproduces the rapid decrease of $R_{4/2}$ with mass number. The
agreement with the empirical ratios is somewhat better in the case of Ba isotopes,
but even for Xe nuclei the differences are not large. The largest deviation from
the experimental value is calculated for $^{130}$Xe: $\approx 0.15$. For the $N=78$
isotones $^{134}$Ba and $^{132}$Xe, the calculated ratios $R_{4/2} \approx 2.3$ are
not very different from the E(5)-symmetry prediction.  An even better agreement with
data is obtained for the calculated isotopic dependence of the transition probabilities
B(E2; $2^+_1 \to 0^+_1$) (panels on the right of Figs.~\ref{R42-Ba} and \ref{R42-Xe}).
For instance, the calculation reproduces in detail
the swift decrease of the empirical B(E2) values from  $\approx 40$
Weisskopf units in $^{128}$Xe, to about 15 Weisskopf units in
$^{134}$Xe. It should be noted that the calculation is completely
parameter-free, i.e. physical observables, such as transition probabilities
and spectroscopic quadrupole moments, are calculated in the full
configuration space and there is no need for effective charges.

\begin{figure}[htb]
\includegraphics[scale=0.4]{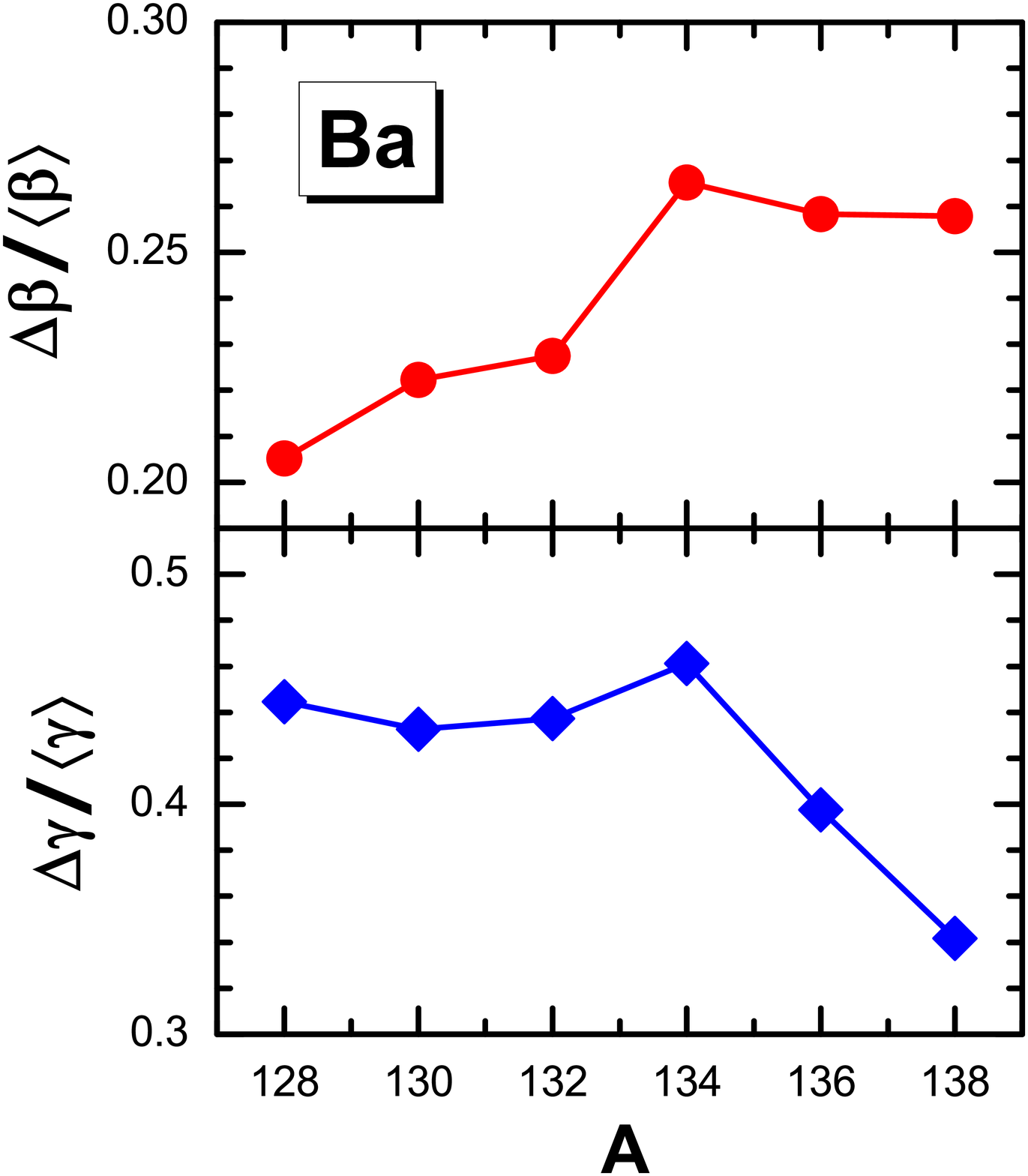}
\caption{\label{fluct-Ba} (Color online) Fluctuations of the quadrupole deformations
${\Delta \beta / <\beta>}$
and ${\Delta \gamma / <\gamma>}$ for the sequence of ground-states of Ba isotopes.}
\end{figure}

\begin{figure}[htb]
\includegraphics[scale=0.4]{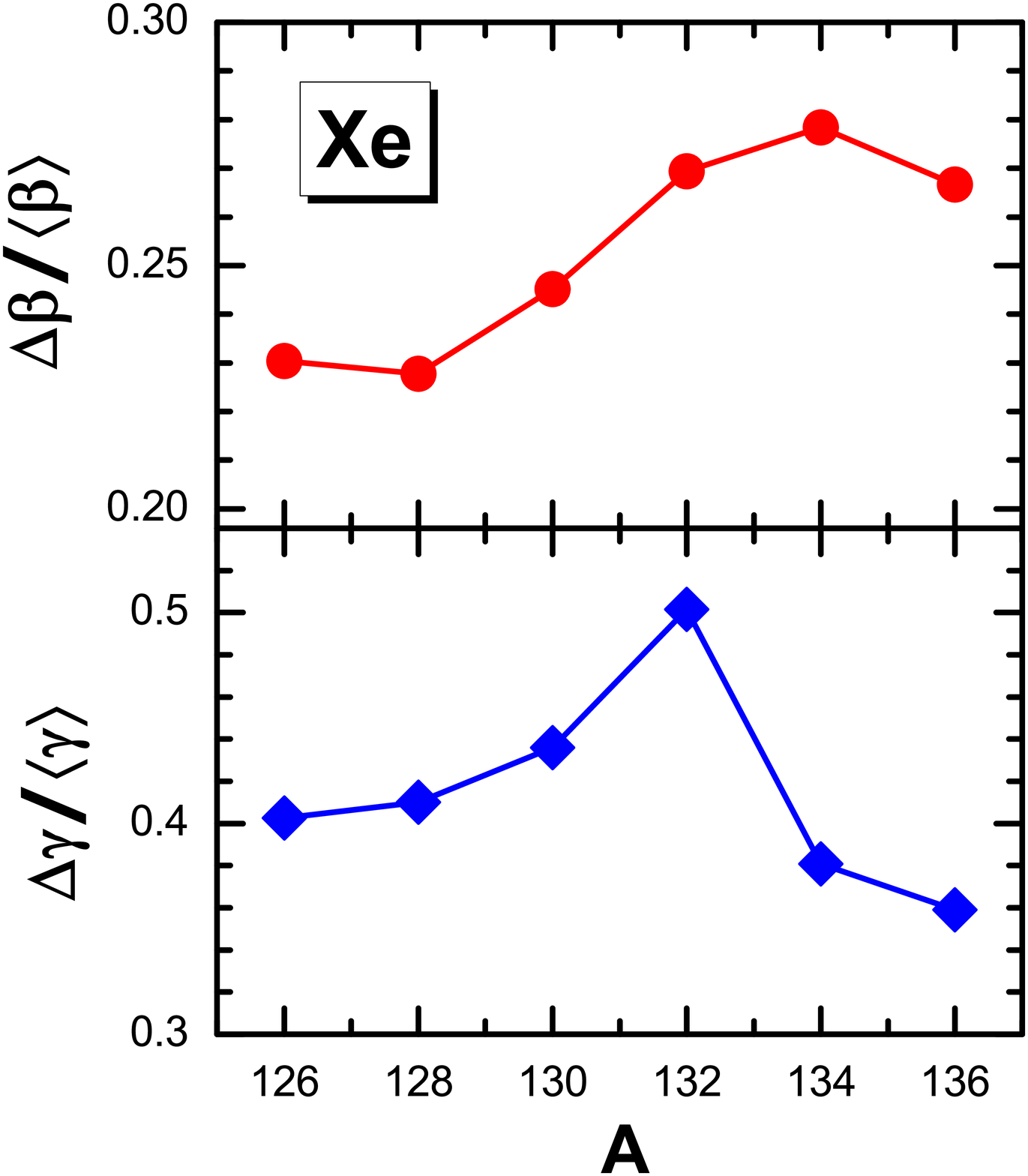}
\caption{\label{fluct-Xe} (Color online) Same as described in the caption to Fig.~\ref{fluct-Ba} but
for the isotopes of Xe.}
\end{figure}

Before comparing the calculated excitation spectra and E2 transition rates with
available data and  E(5)-symmetry model predictions, in Figs.~\ref{fluct-Ba} and \ref{fluct-Xe}
we plot the relative fluctuations of the quadrupole deformations ${\Delta \beta / <\beta>}$
and ${\Delta \gamma / <\gamma>}$, for the sequence of ground-states of Ba and Xe isotopes,
respectively.  The variances are defined by \cite{Srebrny.06}:
\begin{equation}
\Delta \beta = \left.\sqrt{\langle \beta^4\rangle -\langle \beta^2\rangle^2}\right/2<\beta>, \quad
\Delta \gamma = \left.\sqrt{\frac{\langle \beta^6\cos^2{3\gamma}\rangle}{\langle\beta^6\rangle} -
     \frac{\langle\beta^3\cos{3\gamma}\rangle^2}{\langle\beta^4\rangle\langle\beta^2\rangle}} \right/
     3\sin{(3<\gamma>)}\;,
\end{equation}
where the average values of $\beta$ and $\gamma$
\begin{equation}
<\beta> = \sqrt{\langle \beta^2\rangle}, \quad
<\gamma> = \left.\arccos{\left( \left.\langle\beta^3 \cos{3\gamma} \rangle\right/
    \sqrt{\langle\beta^4\rangle\langle\beta^2\rangle} \right)}\right/3\;,
\end{equation}
are calculated in the nuclear ground state, i.e. in the lowest eigenstate of the collective Hamiltonian.
It is interesting to note the change in the fluctuations around $N=78$. For a transition 
from deformed to spherical shapes one expects an increase of the relative fluctuation 
${\Delta \beta / <\beta>}$. For the triaxial deformation variable $\gamma$ the calculation predicts a
marked maximum of the relative fluctuation in $^{134}$Ba and $^{132}$Xe, that can be
interpreted as a possible fingerprint of the second-order shape phase transition.

\begin{figure}[htb]
\includegraphics[scale=0.65,angle=90]{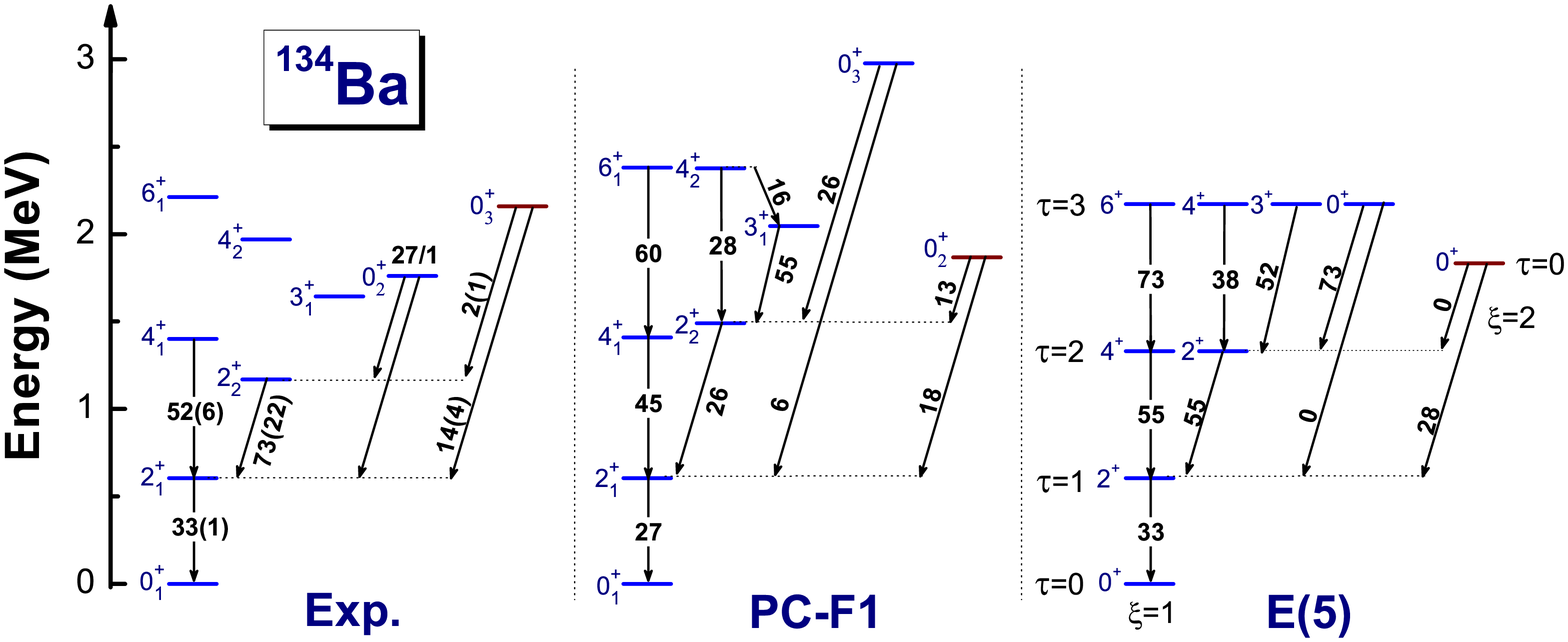}
\caption{\label{spect-134Ba} (Color online) The low-energy
spectrum of $^{134}$Ba calculated with the
PC-F1 relativistic density functional
(middle), compared with the data (left), and the
E(5)-symmetry predictions (right) for the excitation energies, intraband
and interband B(E2) values (in Weisskopf units).
The theoretical spectra are normalized to the experimental energy of the
state $2^+_1$, and the E(5) transition strengths are normalized
to the experimental B(E2; $2^+_1 \to 0^+_1$).}
\end{figure}

$^{134}$Ba was the first nucleus to be considered as a good
example of empirical realization of the E(5) symmetry
at the critical point of second-order phase transition between spherical and
$\gamma$-soft shapes \cite{CZ.00}. In Fig.~\ref{spect-134Ba} we compare the spectrum
of the collective Hamiltonian for $^{134}$Ba with available data for positive
parity states \cite{NNDC,LBNL,Ba134.1,Ba134.2}, and with the predictions of the
E(5) model. For the moments of inertia of the collective Hamiltonian we
have multiplied the Inglis-Belyaev values Eq.~(\ref{Inglis-Belyaev}) with
a common factor determined in such a way that the calculated energy of
the $2_1^+$ state coincides with the experimental value \cite{Ni.09,Li.09}.
This additional scale parameter is necessary because of the well known fact
that the Inglis-Belyaev (IB) formula (\ref{Inglis-Belyaev}) predicts effective
moments of inertia that are considerably smaller than empirical values.
Following the minimal prescription of Ref.~\cite{LGD.99},
the moments of inertia that parameterize the collective Hamiltonian
can be simply related to the IB values:
$\mathcal{I}_k (q) = \mathcal{I}^{IB}_k (q) (1+ \alpha)$, where $q$
denotes the generic deformation parameter, and $\alpha$ is a constant
that can be determined in a comparison with data \cite{Li.09}. 
Below we discuss in more detail the approximations used in the 
calculation of moments of inertia and mass parameters. 

The excitation spectrum of the collective Hamiltonian determined by
the PC-F1 density functional is in good agreement with the available data.
This is especially true for the excitation energies and transitions in the
ground-state band. The calculated sequence of states $2_2^+$,
$3_1^+$, and $4_2^+$, is approximately 300 keV higher than the
corresponding experimental levels, and the transition $2^+_2 \to 2^+_1$
is considerably weaker compared to experiment (notice, however, the
large error bar of the experimental value). By comparing the de-excitation
pattern of the two excited $0^+$ states, it appears that the
ordering of these states is reversed in the theoretical spectrum.
The calculation predicts that the second $0^+$ state predominantly
decays to $2_1^+$, and for the third $0^+$ state a much stronger
transition is predicted to $2_2^+$. In fact, the calculated $0_2^+$ state
is close in energy to the experimental $0_3^+$ state which displays a similar E2 branching 
to $2_1^+$ and $2_2^+$. The third theoretical $0^+$ state, however, is calculated 
more than 1 MeV above the corresponding experimental $0_2^+$ state. One
might also notice that the absolute B(E2) values (in Weisskopf units) for the
stronger transitions are in very good agreement with data, but the calculated
$0_2^+$ and $0_3^+$ states are obviously much more mixed than the
corresponding experimental states. This also explains the strong
repulsion between the theoretical second and third $0^+$ states.

\begin{figure}[htb]
\vspace{3cm}
\hspace{-3cm}
\includegraphics[scale=0.75]{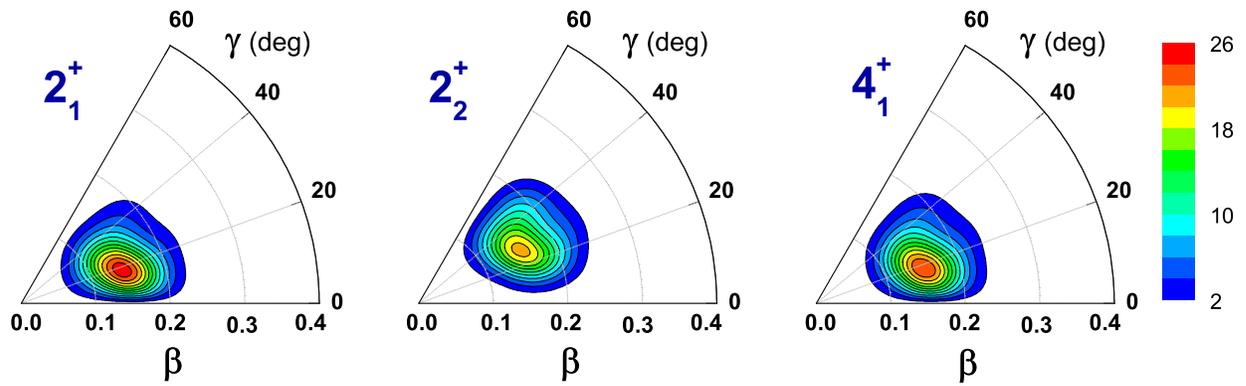}
\vspace{-2cm}
\caption{\label{wavef-134Ba}  (Color online) Probability densities Eq.~(\ref{eq:density})
in the $\beta - \gamma$ plane for the states $2^+_1$, $2^+_2$ and $4^+_1$ of
 $^{134}$Ba. }
\end{figure}

It is interesting to compare the PC-F1 excitation spectrum with the E(5) level scheme
that corresponds to $^{134}$Ba (right panel in Fig.~\ref{spect-134Ba}). In addition
to the total angular momentum $I$, in the case of E(5) symmetry states are labeled
by two quantum numbers: $\xi = 1,2,3,\ldots$ enumerates the zeros of the Bessel functions
$J_{\tau +3/2}$ in $\beta$, and $\tau = 0,1,2,\ldots$ is associated with the $O(5)$ algebra
and denotes the phonon-like levels within a $\xi$-family \cite{FI.00}. The corresponding
spectrum is parameter-free up to an overall scale factor, that is adjusted by normalizing
the energies to the experimental excitation energy of the first excited state $2_1^+$.
The E(5) transition rates are normalized to the experimental B(E2; $2^+_1 \to 0^+_1$).
Some striking similarities are found in the comparison of PC-F1 and E(5) excitation spectra
for $^{134}$Ba. For instance, in the ground state band the calculated ratios
\begin{displaymath}
R_1= { B(E2; 4^+_1 \to 2^+_1) \over B(E2; 2^+_1 \to 0^+_1)} = 1.67\quad\quad
{\rm and} \quad\quad
R_2= { B(E2; 6^+_1 \to 4^+_1) \over B(E2; 4^+_1 \to 2^+_1)} = 1.33\;,
\end{displaymath}
exactly coincide with the values predicted by the E(5) symmetry model.  Again, we emphasize that
the calculation of transition probabilities in the microscopic collective model is parameter-free.
A very good agreement
is also found for B(E2; $4^+_2 \to 2^+_2$) and B(E2; $3^+_1 \to 2^+_2$). The second $0^+$ state
is calculated at an excitation energy very close to the predicted position of  $0^+_{\xi=2,\tau=0}$, 
and this state also predominantly decays to $2_1^+$ ($\Delta \tau =1$ transition in the E(5) model). 
Even though calculated at much higher energy, the decay pattern of the third $0^+$ state is similar 
to that of the third E(5) $0^+$ state with $\xi=1$, and $\tau=3$. The two excited $0^+$ states in the 
E(5) model belong to different $\xi$-multiplets, and only $\Delta \tau =1$ transitions
are allowed. The microscopic collective Hamiltonian, on the other hand, breaks the  
symmetry and this is reflected both in the decay pattern and the excitation
energy of the $0_3^+$ state. Another difference between the PC-F1 and E(5) spectrum is
the transition $2^+_2 \to 2^+_1$, for which the E(5) model predicts the same B(E2) value as
for $4^+_1 \to 2^+_1$. The smaller value calculated from the eigenvectors of the collective
Hamiltonian can be understood by considering the corresponding probability density
distributions. For a given collective state Eq.~(\ref{wave-coll}),
the probability density distribution in the ($\beta,\gamma$) plane is defined by
\begin{equation}
  \label{eq:density}
  \rho_{I\alpha}(\beta,\gamma)=\sum\limits_{K\in\Delta I}
  |\psi^I_{\alpha K}(\beta,\gamma)|^2\beta^3|\sin3\gamma| \;.
\end{equation}
The normalization reads:
\begin{equation}
  \int^\infty_0\beta{\rm d}\beta\int^{2\pi}_0{\rm d}\gamma\ \rho_{I\alpha}(\beta,\gamma)=1\;.
\end{equation}
In Fig.~\ref{wavef-134Ba} we plot the probability densities  in the $\beta - \gamma$ plane
for the states $2^+_1$, $2^+_2$ and $4^+_1$ of $^{134}$Ba. The states $2^+_1$ and
$4^+_1$ belong to the ground-state band, and the corresponding wave functions 
display a large overlap in the $\beta - \gamma$ plane. The average values of $\beta$ are: 
0.156 for $2^+_1$, and 0.166 for $4^+_1$, respectively, and the average values of $\gamma$ 
are: $23.4^0$ for $2^+_1$, and $23.7^0$ for $4^+_1$. Because of the large overlap of the
wave functions, one expects a strong transition between these states. For the state $2^+_2$
the average value  of of $\beta$ is 0.174, but $<\gamma> = 29.6^0$, i.e. the overlap between
the collective wave functions of the states $2^+_1$ and $2^+_2$ is smaller than the one
between $2^+_1$ and $4^+_1$. Consequently, the transition $2^+_2 \to 2^+_1$ is weaker
than $4^+_1 \to 2^+_1$. However, as we have already noted, this result is not in agreement 
with data which show that the transition $2^+_2 \to 2^+_1$ is in fact stronger than 
$4^+_1 \to 2^+_1$.

\begin{figure}[htb]
\includegraphics[scale=0.65,angle=90]{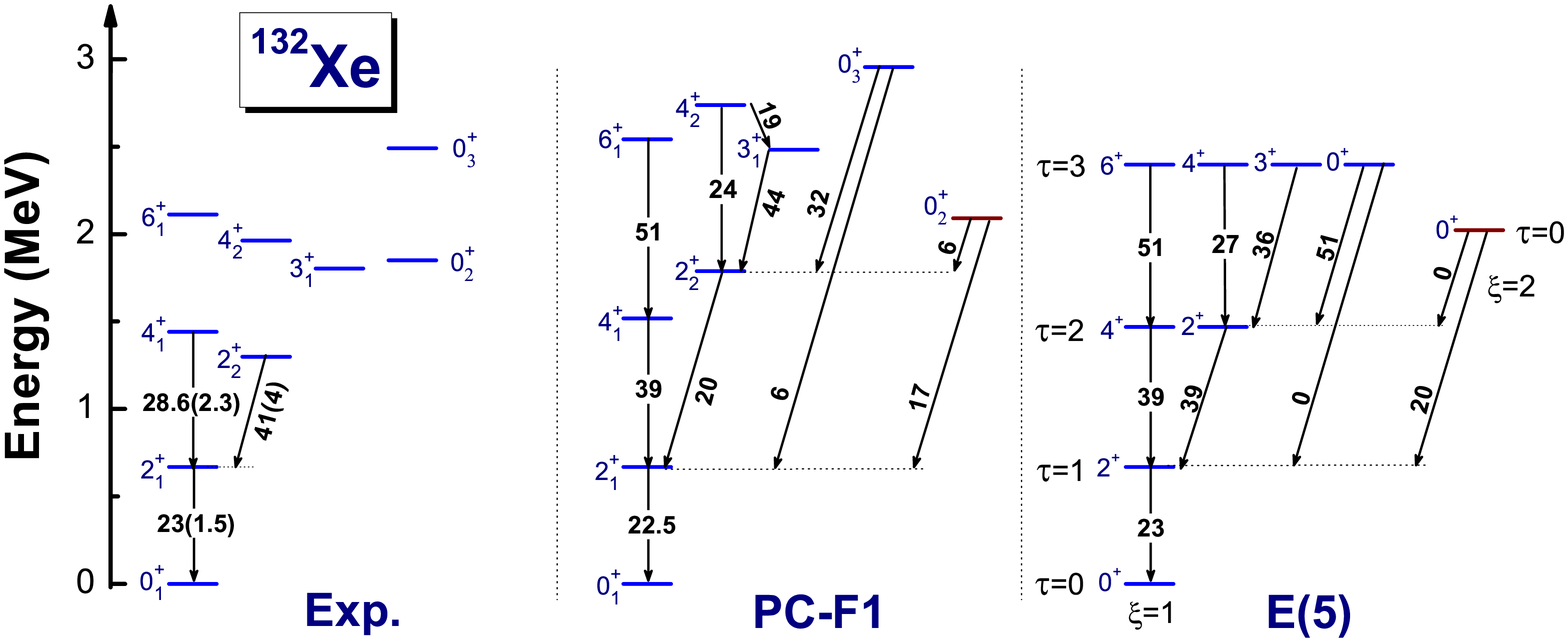}
\caption{\label{spect-132Xe}
Same as described in the caption to Fig.~\ref{spect-134Ba} but for the nucleus $^{132}$Xe.}
\end{figure}

\begin{figure}[htb]
\includegraphics[scale=0.65,angle=90]{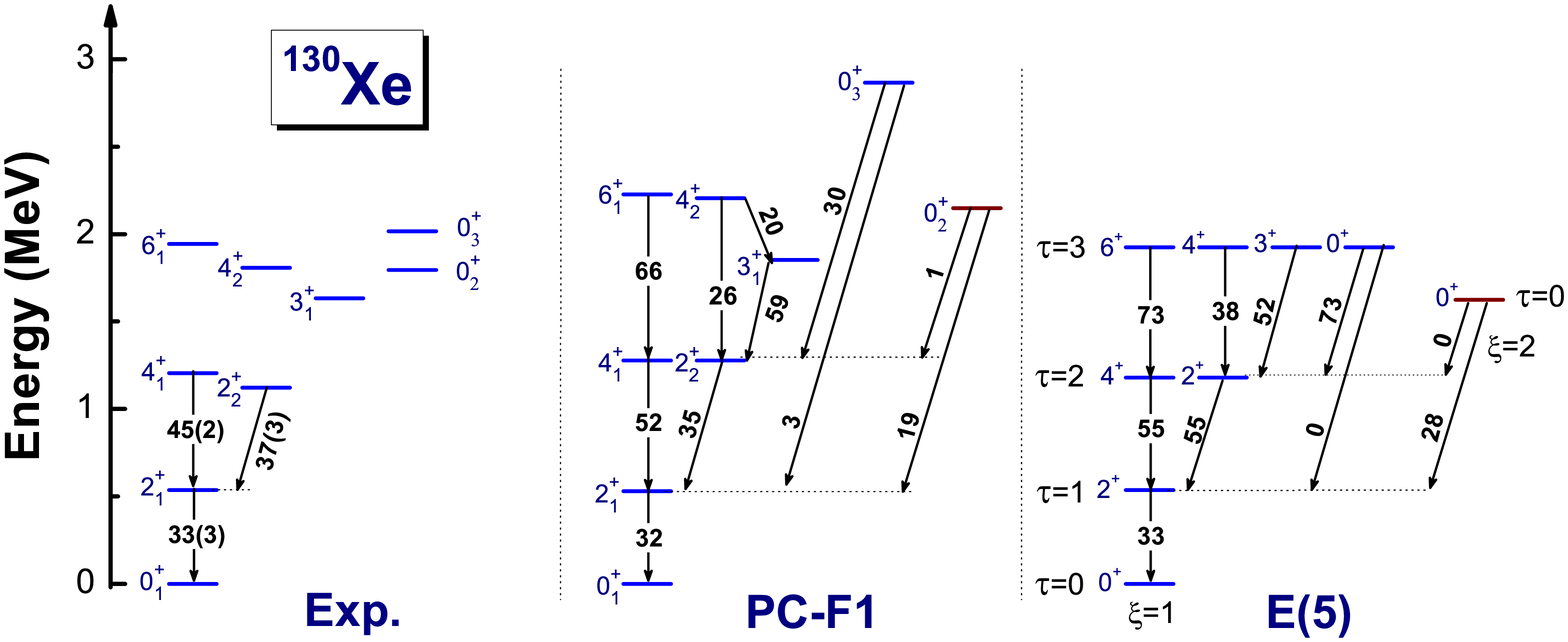}
\caption{\label{spect-130Xe}
Same as described in the caption to Fig.~\ref{spect-134Ba} but for the nucleus $^{130}$Xe.}
\end{figure}

Turning now to the sequence of Xe isotopes, in Fig.~\ref{spect-132Xe} the PC-F1
collective excitation spectrum of $^{132}$Xe, i.e. the isotone of $^{134}$Ba,
is compared with available data \cite{NNDC,Xe130.132}, 
and the level scheme predicted by the E(5) symmetry
model. We find reasonable agreement with experiment and, again, a remarkable similarity
with the decay pattern predicted for the E(5) critical-point symmetry. The limited set of data,
however, do not provide strong evidence for $^{132}$Xe being a good example of 
empirical realization of the E(5) critical-point symmetry. Therefore in Fig.~\ref{spect-130Xe}
we also display the spectra of $^{130}$Xe. In this case the microscopic calculation based on 
the PC-F1 functional predicts an excitation spectrum in slightly better agreement with 
available data  \cite{NNDC,Xe130.132}, e.g. the transitions between low-lying states, 
and the decay scheme is also very similar to the one
that characterizes the E(5) symmetry, except for a weaker transition
$2^+_2 \to 2^+_1$ that has already been discussed in the case of $^{134}$Ba. One might
even notice that the calculated E2 transitions from $0_2^+$ and $0_3^+$ are, for this nucleus,
actually closer to the branching pattern predicted by E(5) than they were in the case of
$^{134}$Ba and $^{132}$Xe. The transitions $0^+_3 \to 2^+_1$ and $0^+_2 \to 2^+_2$,
which are forbidden in the E(5) symmetry limit because of the $\Delta \tau =1$ selection
rule, are strongly suppressed between eigenstates of the PC-F1-based collective Hamiltonian.

\begin{figure}[htb]
\includegraphics[scale=0.65,angle=90]{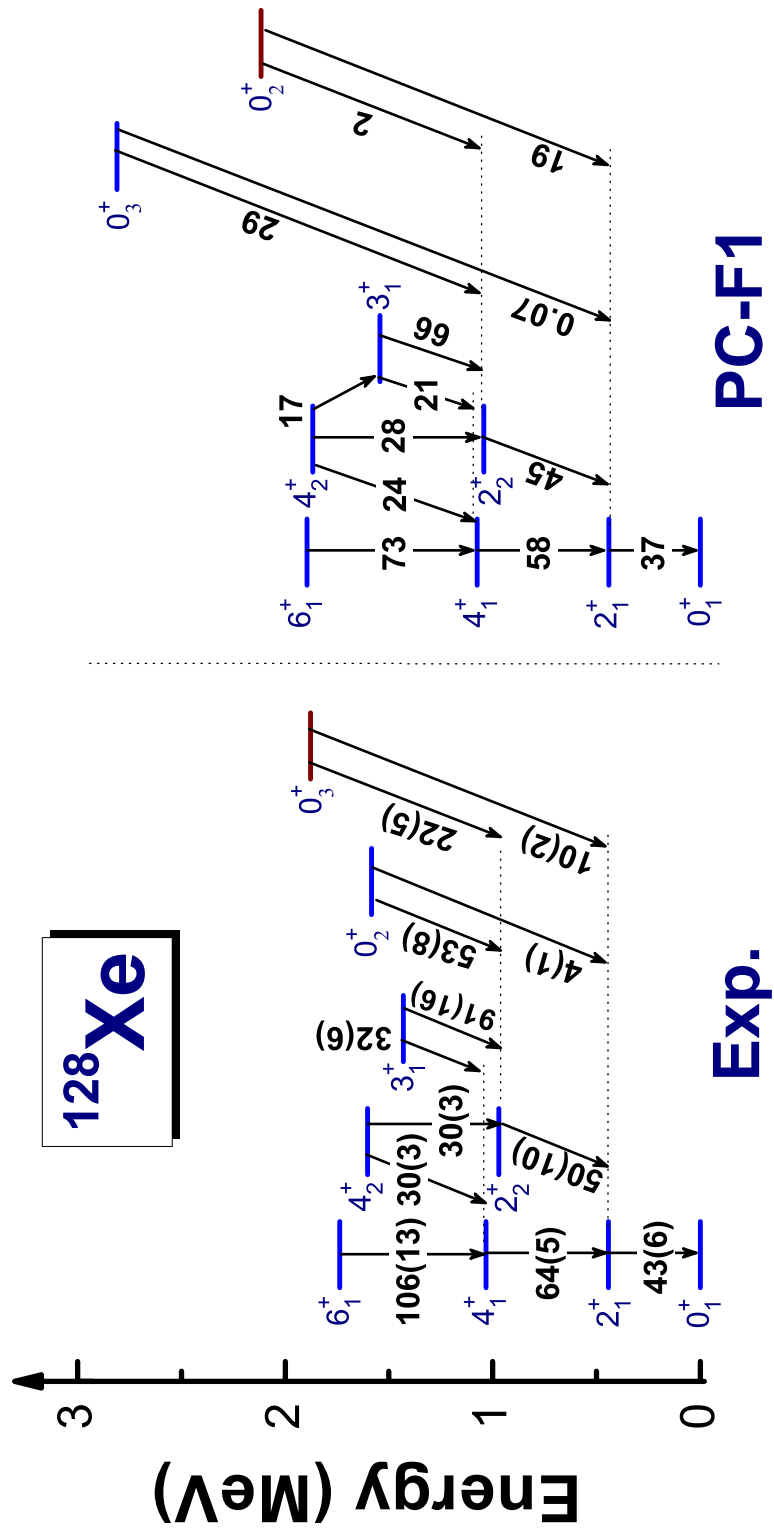}
\caption{\label{spect-128Xe} (Color online) The experimental low-energy
spectrum of $^{128}$Xe (left), compared with the level scheme and
decay pattern predicted by the solution of the microscopic collective
Hamiltonian with the PC-F1 relativistic density functional.}
\end{figure}

In a systematic search for experimental fingerprints of E(5) critical-point symmetry \cite{Clark},
it was suggested that one possible candidate might be $^{128}$Xe. However, for this nucleus
the ratio $R_{4/2} = 2.33$ is intermediate between the value for E(5) ($R_{4/2} = 2.2$) and
the deformed $\gamma$-independent limit or $O(6)$ ($R_{4/2} = 2.5$). In fact, the microscopic
potential of $^{128}$Xe (Fig.~\ref{PES-Xe}) displays a $\gamma$-independent surface for
axial deformations $0.1 \leq \beta \leq 0.2$. We also notice that the potential is not nearly
as rigid in $\beta$ for $\beta \leq 0.1$ and $\beta > 0.2$, as one assumes in the E(5) symmetry limit.
In Ref.~\cite{Bon.06} a $\gamma$-soft analog of the confined $\beta$-soft (CBS) rotor
model \cite{PG.04} has been constructed, that corresponds to an infinite square-well
potential over a confined range of values $\beta_M > \beta_m \geq 0$
(confined $\beta$-soft potential, $\beta_m$ and $\beta_M$ denote the positions of the inner
and outer walls on the $\beta$ axis, respectively). The $\gamma$-soft extension of Ref.~\cite{Bon.06}
generalizes the E(5) critical-point symmetry to a parametric solution over the whole path between E(5)
and the $\beta$-rigid deformed $\gamma$-independent limit. The structure of $^{128}$Xe was
investigated and, from the decay pattern of the first two excited $0^+$ states, it was suggested that
$^{128}$Xe is located well beyond the $U(5) \to O(6)$ shape-phase transition in the deformed
phase. In a very recent experimental study of low-lying collective states in $^{128}$Xe by
$\gamma$-ray spectroscopy \cite{128Xe}, the data, including the first measurement of the
B(E2) values for transitions from the $0_2^+$ and $0_3^+$ states, were compared with
theoretical predictions and it was concluded that $^{128}$Xe is not an E(5) nucleus, thus
leaving $^{130}$Xe as the best candidates among the Xe isotopes. In Fig.~\ref{spect-128Xe}
we, therefore, finally compare our microscopic results for $^{128}$Xe with the recent experimental
decay scheme of Ref.~\cite{128Xe}. The calculated excitation energies and B(E2) values for
the ground-state band and the sequence $2^+_2$, $3^+_1$, $4^+_2$, are in good agreement
with experiment. The two excited $0^+$ states, however, are calculated at higher energy,
especially $0^+_3$. The predicted transitions to $2^+_1$ and $2^+_2$ clearly follow the
pattern calculated in $^{132}$Xe and $^{130}$Xe, and do not reproduce the empirical two-level
mixing that strongly breaks the $\Delta \tau = 1$ selection rule characteristic for $O(5)$ symmetry.

For all nuclei analyzed in this work we have found a very good agreement between data and
the predictions of the microscopic collective Hamiltonian based on the PC-F1 energy density
functional plus $\delta$-force pairing, except for the two excited $0^+$ states. In particular, the 
second excited $0^+$ state, i.e. the one that predominantly 
decays to $2^+_2$,  is systematically calculated at
considerably higher excitation energy compared to the experimental position of the two
excited $0^+$ levels. In $^{134}$Ba and $^{128}$Xe, for which data on absolute E2 transition
strengths are available, it is the first excited $0^+$ state that predominantly decays to $2^+_2$,
contrary to the results of our calculation. One reason for this discrepancy, of course, could be
the particular choice of the energy density functional and/or the treatment of pairing correlations,
but also the approximation used in the calculation of the mass parameters (vibrational inertial
functions). In the current version of the model the moments of inertia are calculated from the
Inglis-Belyaev formula (Eq.~(\ref{Inglis-Belyaev})). These moments of inertia do not include the
Thouless-Valatin dynamical rearrangement contributions and, therefore, are systematically smaller
than the empirical values. However, as shown in several studies \cite{LGD.99}, the Thouless-Valatin
corrections are almost independent of deformation, and the effective moments of inertia to be
used in the collective Hamiltonian can simply be obtained by renormalizing the Inglis-Belyaev
values by a constant factor, characteristic for a given nucleus. The situation is considerably more
complicated in the case of mass parameters \cite{Pro.04,Pro.09}, for which there are no simple
estimates of the Thouless-Valatin correction, especially for nuclei with $\gamma$-soft potential
energy surfaces. Some authors \cite{Pro.09} argue that, to approximately take into account the
Thouless-Valatin correction, all inertial functions, not only the moments of inertia, should be
rescaled by a constant multiplicative factor. 

Another effect that is not included in our model is the coupling of nuclear shape oscillations 
with pairing vibrations, i.e. vibrations of the pairing density. A number of studies have shown 
that excited $0^+$ states are very sensitive to the coupling between these modes. For the 
Ba and Xe isotopes, in particular, the effect of coupling between pairing vibrations and 
axial quadrupole vibrations was investigated in Ref.~\cite{PP.93}, using collective Hamiltonians 
based on a microscopic Nilsson single-particle Hamiltonian with pairing forces and a long-range 
residual interaction.  It was shown that the coupling between the pairing and quadrupole 
modes has a very pronounced effect on the lowest excited $0^+$ states, lowering their excitation 
energies by $40 - 50\%$, in much closer agreement to data. In an even earlier study of collective 
states in Xe and Ba isotopes \cite{Roh.77}, based on the general 5-dimensional Bohr Hamiltonian, 
this coupling was not taken into account explicitly, but rather it was simulated by a reduction of the 
pairing strengths of $\approx 20\%$. Even though the resulting pairing interaction is somewhat  
unphysical, i.e. it gives proton and neutron gaps that are too small, the resulting 
excitation energies of the lowest $0^+$ states are $40 - 50\%$ lower, in better agreement 
with experiment. 

The present study, however, employs exactly the same model that was used in 
our previous analysis of spherical to axially-deformed shape
transitions, The mass parameters are calculated in the cranking approximation
Eqs.~(\ref{masspar-B}) and (\ref{masspar-M}), and from these expressions it is obvious that the
resulting mass parameters will be very sensitive to the underlying microscopic shell structure and
to the treatment of pairing correlations. In fact, we have found that the position of the first excited
$0^+$ state in our calculation (the one that predominantly decays to $2^+_1$) is sensitive to
the value of the mass parameter $B_{\beta\beta}$, the excitation energy of the third $0^+$ state
(predominantly decaying to  $2^+_2$) displays a very strong dependence on $B_{\gamma\gamma}$,
and the mixing of the two excited $0^+$ states is controlled by $B_{\beta\gamma}$. For instance,
by decreasing $B_{\beta\gamma}$, i.e. by reducing the coupling between $\beta$ and $\gamma$
vibrational degrees of freedom, the mixing between $0^+_2$ and $0^+_3$ decreases and the
resulting spectrum is even closer to the E(5) level scheme. By increasing the value of
$B_{\gamma\gamma}$ one finds a rapid decrease of the excited energy of the calculated
$0^+_3$, and eventually the two excited $0^+$ state cross, in better agreement with available
data for $^{134}$Ba and $^{128}$Xe. The only other effect on the calculated spectrum is a
much smaller lowering of the sequence: $2^+_2$, $3^+_1$, $4^+_2$, again in agreement with 
experiment. In the case of $^{128}$Xe,
for instance, an enhancement of $B_{\gamma\gamma}$ by a factor $\geq 1.5$ brings the calculated
excitation energies in very good agreement with the experimental spectrum shown in
Fig.~\ref{spect-128Xe} \cite{128Xe}.  

Very recently a global study of low-energy nuclear structure at normal deformation has 
been reported, based on the non-relativistic Hartree-Fock-Bogoliubov framework with the 
Gogny interaction D1S, and mapped onto a five-dimensional collective quadrupole
Hamiltonian \cite{Del.10}. Ground-state properties, excitation energies, quadrupole and 
monopole transition matrix elements for yrast levels up to $6^+_1$ and the lowest 
yrare levels, have been calculated for even-even particle-stable nuclei with proton 
numbers $Z=10$ to $Z=110$ and neutron number $N\leq 200$. Except for the use of 
a different effective interaction in the particle-hole and particle-particle channels, 
the model of Ref.~\cite{Del.10} is very similar to the one employed in the present study. 
 However, while in both models the quadrupole mass parameters are calculated in the 
 cranking approximations, in the calculation of the moment of inertia 
 the model of Ref.~\cite{Del.10} goes beyond the simple Inglis-Belyaev formula used in 
 this study, and explicitly includes the Thouless-Valatin dynamical rearrangement contributions. 
 This leads to an increase of the moments of inertia and the resulting excitation energies of 
 yrast states in deformed and transitional nuclei are, in general, in very good agreement with 
 data. As a particular example, the study of Ref.~\cite{Del.10} has considered the low-energy 
 collective quadrupole spectrum of $^{152}$Sm. It has been shown that the spectrum 
 calculated with the collective Hamiltonian based on the Gogny interaction D1S is 
 consistent with the X(5) analytical solution at the critical point of QPT between spherical 
 and axially deformed shapes in $N \approx 90$ isotones.  More generally, the calculated 
 spectral properties of $\approx 1700$ even-even nuclei have demonstrated the predictive 
 power of the quadrupole collective Hamiltonian model based on a microscopic effective 
 interaction. A problem common to both studies, i.e. to the global survey of Ref.~\cite{Del.10} 
 and the specific calculation of transitional Ba and Xe isotopes reported in this work, are the 
excitation energies of excited $0^+$ states that are systematically calculated too high compared 
with data. Similar to the discussion above, in Ref.~\cite{Del.10} it has been suggested 
that the energies of excited $0^+$ states could be improved by including the 
Thouless-Valatin dynamical rearrangement contributions in the calculation of the 
quadrupole mass parameters, and by extending the model space to include 
two-quasiparticle components in the wave functions.

\section{\label{secIV}Conclusions}
The recently developed implementation for the solution of the five-dimensional
collective Hamiltonian for quadrupole vibrational and rotational degrees of freedom,
with parameters determined by constrained self-consistent relativistic mean-field
calculations for triaxial shapes, has been applied to the study of
$\gamma$-soft shapes in Ba and Xe nuclei in the mass $A \geq 130$ region.
This microscopic approach had previously been used to describe
spherical to axially deformed shape transitions in the mass $A \approx 150$
region, and the results have shown there is an abrupt change of structure at
$N=90$ that can approximately be characterized by the X(5) analytic solution
at the critical point of a first-order quantum phase transition. In this sense
the present analysis represents a natural extension of our previous studies
\cite{Ni.07,Li.09,Li.09a}, to a region of nuclei where evidence has been reported for
a second-order QPT between spherical and $\gamma$-soft shapes, with the
critical point of phase transition characterized by the E(5) dynamical symmetry.
Without any modification of the model Hamiltonian, of the treatment of pairing
correlations, or the parameters of the effective interactions in the particle-hole
and particle-particle channels, we have been able to describe the rapid
transition between spherical and $\gamma$-soft shapes in Ba and Xe isotopes.
The results reproduce the characteristic evolution of excitation spectra and E2
transition probabilities, and a good agreement with available data is obtained.
In specific nuclei, e.g. $^{134}$Ba and $^{132}$Xe, the calculated triaxial quadrupole
binding energy surfaces are rather flat in an extended interval of the axial deformation
parameter $\beta$, and almost independent of the $\gamma$-deformation. For the
ground states of these nuclei one finds pronounced discontinuities in the relative
fluctuations of the quadrupole deformations ${\Delta \beta / <\beta>}$
and ${\Delta \gamma / <\gamma>}$. The calculated spectra are remarkably
similar to the level schemes predicted by the E(5) dynamical symmetry model and,
therefore, display fingerprints of a second-order shape phase transition.

The fact that the same microscopic model describes both first- and second-order
QPT in different mass regions is, in our opinion, an interesting result which illustrates
the importance of microscopic studies of QPT starting from nucleonic degrees of
freedom. A quantitative comparison with experimental spectra, particularly with very
recent data for $^{128}$Xe, points to some intrinsic problems in the description
of excited $0^+$ states and clearly indicates that one must go beyond the simple
cranking approximation for the model mass parameters (vibrational inertial
functions), or even extend the model by explicitly taking into account the 
coupling of nuclear shape oscillations with pairing vibrations.

\begin{acknowledgements}
We thank R. F. Casten, F. Iachello, and P. Ring for useful discussions.
This work was supported in part by MZOS - project 1191005-1010,
by the Major State 973 Program 2007CB815000, and
the NSFC under Grant Nos.10775004 and 10975008.
The work of J.M, T.N., and D.V. was supported in part by
the Chinese-Croatian project "Nuclear structure and astrophysical applications".
T. N. and Z. P. Li acknowledge support by the Croatian National Foundation for
Science, Higher Education and Technological Development.
\end{acknowledgements}


\clearpage
\clearpage
\clearpage
\clearpage
\clearpage
\clearpage
\clearpage
\clearpage
\clearpage
\clearpage
\clearpage
\clearpage
\clearpage

\end{document}